\documentstyle[newarcrc,fleqn]{article}

\input{psfig.sty}

\title{A 6.4-hr positive superhump period in TV Col}

\author{Retter A., Hellier C. \address{Department of Astronomy, Keele 
University, Staffs., ST5 5BG, UK; \\ ar@astro.keele.ac.uk; 
ch@astro.keele.ac.uk}}

\begin{document}
\maketitle




\section{Introduction}

So far four periods have been discovered in the light curve of TV Col
(Hellier 1993). They were interpreted as follows:
the 32-min -- the spin period; 
the 5.5-hr period -- the orbital binary revolution; 
the 4-day period -- the nodal precession of the accretion disc,
and the 5.2-hr period -- the beat between the two longer periods 
(a negative superhump). 
This interpretation makes TV Col the permanent superhump system with 
the largest orbital period. Since light curves of many permanent 
superhumpers show both types of superhumps (Patterson 1999), we decided 
to search for positive superhumps in the light curve of TV Col as well.
Extrapolating the Stolz \& Schoembs (1984) relation we predicted that the
superhump period should be around 6.4 hr. Indeed we found such a periodicity 
in the data.

\section{Discussion}

We re-examined four sets of photometric data of TV Col (Hellier 
1993). Three sets show a similar pattern. In the upper panel of Fig. 1 we 
present the power spectrum of the 1989 January run. There is a 
triple alias structure around the three marked peaks. Two of them are the 
known 5.2-hr and 5.5-hr periods. The third peak corresponds to 
the period 6.4 hr. As a first test, we fitted the two known periods 
to the data, subtracted them and performed a new power spectrum on the 
residuals, which is shown in the lower panel of Fig. 1. The third peak 
didn't disappear from the synthetic power spectrum, but even gained power. 


\begin{figure}

\psfig{figure=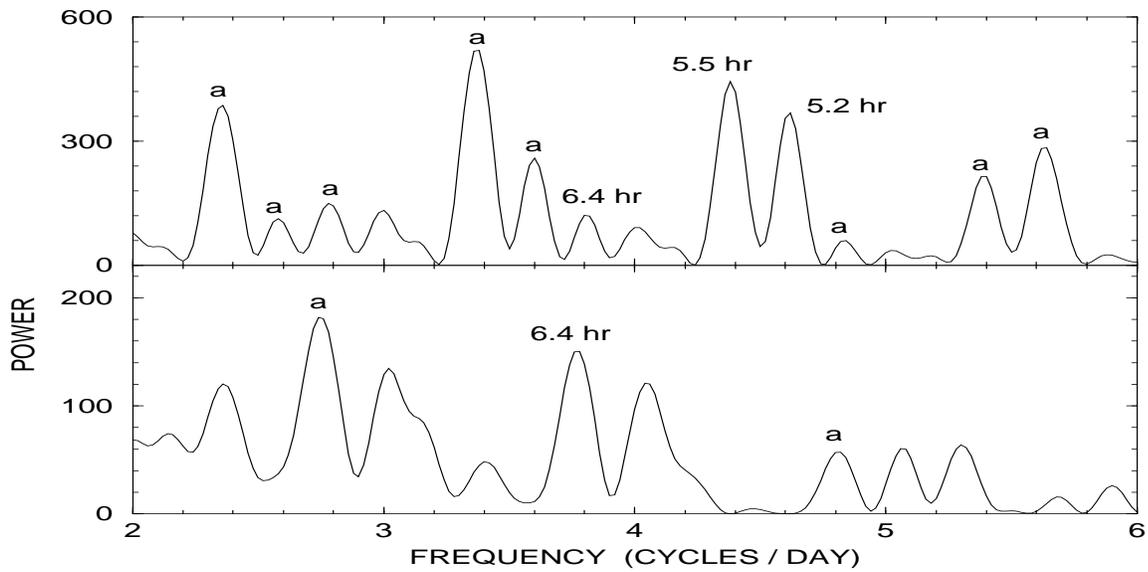,width=6.0in,height=3.0in}

\caption{Upper panel -- power spectrum of the nights of 1989 January.
The `a' signs denote 1-day aliases of the three main peaks.
Lower panel -- the power spectrum after the removal of the two known periods.
The difference between the marked 6.4-hr peak and the one just at its right 
is nearly the nodal precession frequency.}

\end{figure}



The third period is 0.265+/--0.005 day -- about 16 percent longer than the 
orbital period. It obeys the relation between superhump-period excess and 
orbital period (Stolz \& Schoembs 1984), which is plotted in Fig 2. 
A positive superhump interpretation is inevitable.


\begin{figure}

\psfig{figure=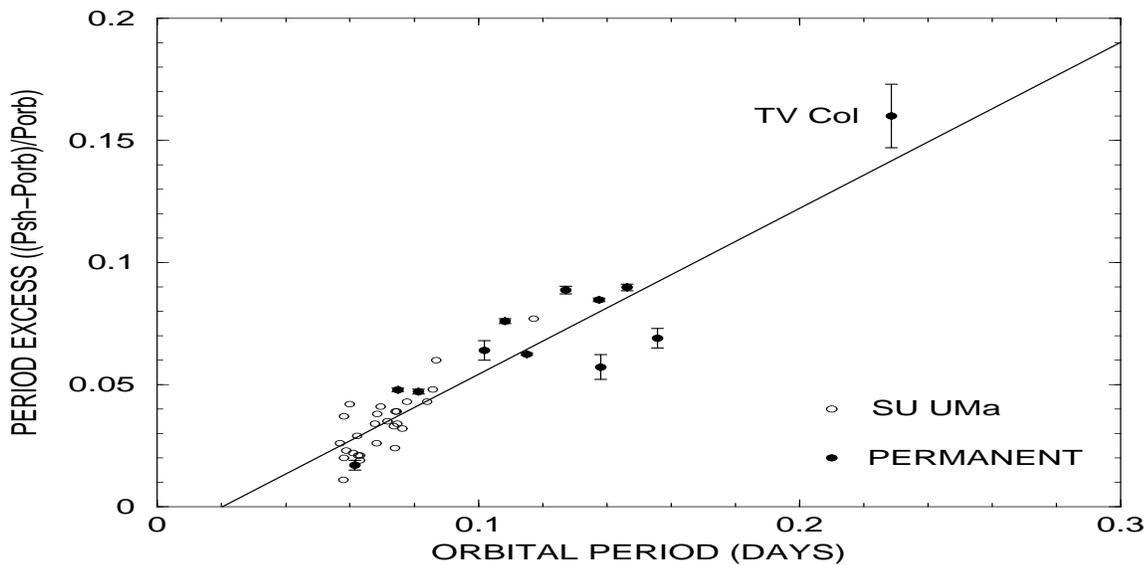,width=6.0in,height=3.0in}

\caption{Stolz \& Schoembs (1984) relation for superhump periods. The new
period of TV Col obeys the trend.}

\end{figure}

As a confirmed permanent superhumper, the accretion disc of TV Col is
naturally thermally stable. Therefore, our result supports the idea of
Hellier \& Buckley (1993) that the short-term outbursts seen in its light
curve are mass transfer events rather than thermal instabilities in the disc.

At 5.5-hr, TV Col has an orbital period longer than any known superhumper, 
and thus a mass ratio which is probably outside the range at which 
superhumps can occur according to the current theory.

\end{document}